%% file: m2actSrcs.tex
\documentclass[twocolumn]{aastex631} 
\graphicspath{{./}{figures/}{figures/final_source_plots/}{figures/new_SEDs/}}

\begin{document}

\title{Sensitive 3~mm Imaging of Discrete Sources in the Fields of Thermal Sunyaev–Zel’dovich Effect-Selected Galaxy Clusters}

\correspondingauthor{Simon R Dicker}
\email{sdicker@hep.upenn.edu}
\author[0000-0002-1940-4289]{Simon R Dicker}
\affiliation{Department of Physics and Astronomy, University of Pennsylvania, 209 S. 33rd St., Philadelphia, PA 19014, USA}
\author{Karen Perez Sarmiento}
\affiliation{Department of Physics and Astronomy, University of Pennsylvania, 209 S. 33rd St., Philadelphia, PA 19014, USA}
\author[0000-0002-8472-836X]{Brian Mason}
\affiliation{National Radio Astronomy Observatory, 520 Edgemont Rd., Charlottesville, VA 22901, USA}
\author[0000-0002-2971-1776]{Tanay Bhandarkar}  
\affiliation{Department of Physics and Astronomy, University of Pennsylvania, 209 S. 33rd St., Philadelphia, PA 19014, USA}
\author{Mark J. Devlin}
\affiliation{Department of Physics and Astronomy, University of Pennsylvania, 209 S. 33rd St., Philadelphia, PA 19014, USA}
\author[0000-0003-3586-4485]{Luca Di Mascolo}
\affiliation{Laboratoire Lagrange, Université Côte d’Azur, Observatoire de la Côte d’Azur, CNRS, Blvd de l’Observatoire, CS 34229, 06304 Nice cedex 4, France}
\author{Saianeesh Haridas}
\affiliation{Department of Physics and Astronomy, University of Pennsylvania, 209 S. 33rd St., Philadelphia, PA 19014, USA}
\author[0000-0002-8490-8117]{Matt Hilton}
\affiliation{Wits Centre for Astrophysics, School of Physics, University of the Witwatersrand, Private Bag 3, 2050, Johannesburg, South Africa}
\affiliation{Astrophysics Research Centre, School of Mathematics, Statistics, and Computer Science, University of KwaZulu-Natal, Westville Campus, Durban 4041, South Africa}
\author{Mathew Madhavacheril}
\affiliation{Department of Physics and Astronomy, University of Pennsylvania, 209 S. 33rd St., Philadelphia, PA 19014, USA}
\author[0000-0001-9793-5416]{Emily Moravec}
\affiliation{Green Bank Observatory, P.O. Box 2, Green Bank, WV 24944, USA}
\author[0000-0003-3816-5372]{Tony Mroczkowski}
\affiliation{ESO -- European Southern Observatory, Karl-Schwarzschild-Str.\ 2, D-85748 Garching b.\ M\"unchen, Germany}
\author[0000-0003-1842-8104]{John Orlowski-Scherer}
\affiliation{Department of Physics and Astronomy, University of Pennsylvania, 209 S. 33rd St., Philadelphia, PA 19014, USA}
\author[0000-0001-5725-0359]{Charles Romero}
\affiliation{Center for Astrophysics $\vert$ Harvard \& Smithsonian, 60 Garden Street, Cambridge, MA 02138, USA}
\author{Craig L. Sarazin}
\affiliation{Department of Astronomy, University of Virginia, 530 McCormick Road, Charlottesville, VA 22904-4325, USA}
\author{Jonathan Sievers}
\affiliation{Department of Physics, McGill University, 3600 University Street Montreal, QC, H3A 2T8, Canada}



\begin{abstract}
In this paper, we present the results of a blind survey for compact sources in 243 galaxy clusters that were identified using the thermal Sunyaev-Zel'dovich effect (tSZ).  The survey was carried out at 90~GHz using MUSTANG2 on the Green Bank Telescope and achieved a $5\sigma$ detection limit of 1~mJy in the center of each cluster. 
We detected 24 discrete sources. The majority (18) of these correspond to known radio sources, and of these, five show signs of significant variability, either with time or in spectral index. The remaining sources have no clear counterparts at other wavelengths.
Searches for galaxy clusters via the tSZ effect strongly rely on observations at 90~GHz, and the sources found have the potential to bias mass estimates of clusters.  We compare our results to the Websky simulation that can be used to estimate the source contamination in galaxy cluster catalogs. While the  simulation showed a good match to our observations at the clusters' centers, it does not match our source distribution further out.  Sources over 104$''$ from a cluster's center bias the tSZ signal high, for some of the sources found, by over 50\%.  When averaged over the whole cluster population, the effect is smaller but still at a level of 1 to 2\%.    We also discovered that unlike previous measurements and simulations, we see an enhancement of source counts in the outer regions of the clusters and fewer sources than expected in the centers of this tSZ-selected sample. 

\end{abstract}


\section{Introduction} \label{sec:intro}

Galaxy clusters are the most massive virialized structures in the Universe, and  they constitute extremely useful laboratories for probing cosmology. For example, cluster counts have been used to estimate cosmological parameters
\citep[e.g.,][]{Komatsu2011} and can help discern between different models for the nature of dark energy \citep{Allen_et_al_2011}. 
Optical and X-ray surveys have long been used to detect and study galaxy clusters. More recently, millimeter-wave surveys have been used to find clusters by the  thermal Sunyaev-Zel'dovich effect \citep[tSZ, see][for a review]{MroczkowskiSZReview}.
The tSZ is a spectral distortion of the Cosmic Microwave background (CMB) due to inverse Compton scattering off hot electrons in the intracluster medium (ICM).  As this is nearly redshift-independent, the tSZ is a powerful tool for detecting galaxy clusters.  Experiments such as the Atacama Cosmology Telescope \citep[ACT;][]{ACT} and the South Pole Telescope \citep[SPT;][]{sptref} have discovered many clusters.  The latest catalogs from the SPT contain over 600 optically confirmed clusters \citep{bleem2020,bleem2023}, while Data Release 5 (DR5) from ACT contains over 4000 \citep{Hilton2020}. In the future, the Simons Observatory and CMB-S4 are expected to discover an order of magnitude more clusters \citep{SimonsForecastPaper,cmb-s4paper}.

\subsection{Clusters and Cosmology}
In order to be used as cosmological tools, it is necessary to know each cluster's mass.   The magnitude of the tSZ effect, the Compton-$y$ parameter, is proportional to the integrated line-of-sight pressure, and its sum over the cluster, $Y$, is a good proxy for the cluster mass. For large surveys, the sum over the cluster is typically calculated using multifrequency matched filters normalized such that the peak at the cluster location is equal to $Y$. This is the centralized Compton-$y$ parameter and in the ACT DR5 cluster catalog is refereed to as $y_c$ \citep{Hilton2020}. There is however, an 11\% scatter in the ACT Compton-$y$ to mass relation \citep[$Y$-$M$;][]{Battaglia2012}.  For tSZ galaxy cluster surveys to fulfill their promise as precision cosmology tools, it is crucial to understand these systematic effects. The mass selection effects in tSZ galaxy cluster surveys 
are well understood and simulations have been used to predict the effects of cluster mergers \citep{MergersAndSZE}. However, past studies disagree on the effect that compact radio sources have on the detection of galaxy clusters and how much this might contribute to the scatter in $Y$-$M$ relationships \citep{Coble2007,Gralla2020}. 

Although the mass of clusters can be found by weak lensing, the weak-lensing signal is not detectable in a single cluster. It is necessary to stack clusters with similar masses (or measured Y values). Scatter in measured Y has the potential to bias weak lensing measurements as some clusters in each bin could be of significantly different mass.   

\citet[hereafter D2021]{Dicker2021} used an ad hoc sample of galaxy clusters observed at 90~GHz by MUSTANG2 on the Green Bank Telescope (GBT) to investigate the effects of compact ($<15''$ in diameter) sources on the measured $y_c$.  This sample of 30 clusters had been selected for deep observations by different projects for a range of scientific aims.  Approximately one third of the clusters had compact sources brighter than 1~mJy, enough to bias individual mass estimates by 2\%--100\%.  However, these clusters were selected for study due to their interesting properties (e.g. known mergers), which could bias the results.  This motivated the study in this paper of a search for sources in a more representative sample of clusters.

\subsection{Compact sources}
\begin{figure}
    \centering
   \includegraphics[width=0.95\linewidth]{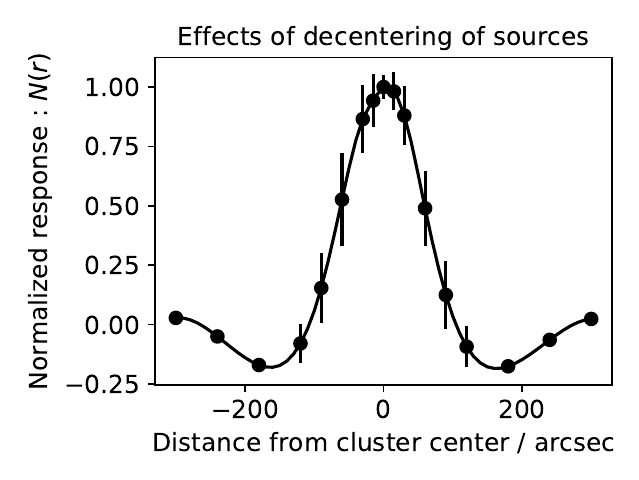}
   \includegraphics[width=0.95\linewidth]{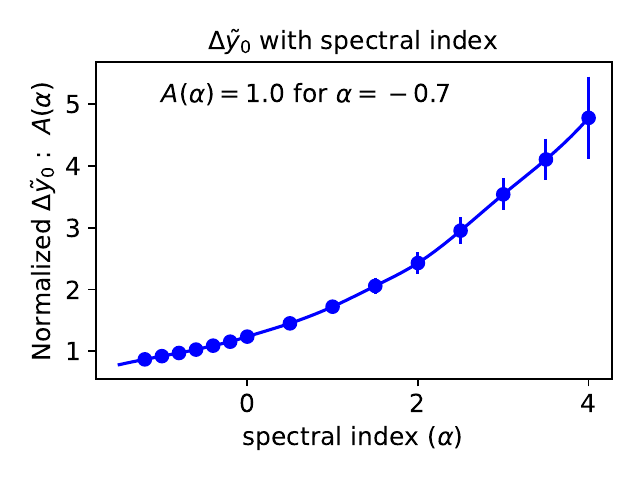}
   \caption{How a source of fixed amplitude at 90~GHz affects the recovered Compton-$y$ parameter varies depending on its projected distance from the cluster center and the source spectral index.  [Reproduced from \citet{Dicker2021}].  This can be represented by the functions
   $N(r)$ (upper panel) which has a null at 104$''$ 
   and $A(\alpha)$ (lower panel). 
   The exact shape depends on the matched filters used by a given tSZ survey but will be similar to these examples from ACT DR5. }
              \label{fig:flux2yc}%
\end{figure}

To maximize mapping speed, tSZ survey instruments have a beam size comparable to the angular size of clusters ($1.'4$--$2.'2$ for ACT).  At these resolutions, compact sources can blend in with the cluster. The sources can be contained within the cluster or be in the foreground or background.  Most of the sensitivity of tSZ measurements comes from a combination of data taken in the 90 and 150~GHz atmospheric windows where the tSZ signal is negative. As shown in Figure~\ref{fig:flux2yc}, a source with a projected location near the cluster center cancels out some of this signal resulting in a decrease in measured cluster mass. When a source has a line of sight location further from the cluster center, the matched filters in the data pipelines used to find clusters can result in an increase in the measured mass of a cluster 
(D2021). The magnitude of the effect depends on the flux of the source at each tSZ band (90 and 150~GHz) which can also be expressed as a flux at 90~GHz and a spectral index $\alpha$ between the two.  These spectral indexes can vary widely depending on source type and are often different from spectral indexes calculated at lower frequencies.  Following the notation of D2021, the effect of a source with a 90~GHz flux $I$ on the measured centralized ACT Compton-$y$ ($y_c$) is given by:
\begin{equation}\label{equ:dy0}
\Delta_{\tilde{y}_0} = I \:\delta_{\tilde{y}_0}\: N(r)\: A(\alpha)
\end{equation}
where $\delta_{\tilde{y}_0}= -8.76{\times}10^{-6}$ is a constant equal to the effect on the Compton-$y$ of a 1~mJy source at 90 GHz, $r=0$, and with a spectral index of $-0.7$,  and $N(r)$ and $A(\alpha)$ are functions representing the response to a source as a function of projected radius ($r$) from the cluster center and spectral index ($\alpha$).

This relationship, found by passing clusters with simulated sources through the real ACT DR5 cluster pipeline,\footnote{\url{https://nemo-sz.readthedocs.io/en/latest/}} can be used to marginalize over the effects of compact sources --- but only if the distribution in 90~GHz flux, the distribution in spectral index and the typical locations within a cluster are known.  Much of our information on compact sources in clusters comes from low-frequency (1--4~GHz) radio surveys. To predict the effect of compact sources on cluster mass determination would therefore require flux extrapolations over 1--2 orders of magnitude.  In addition, near to 90~GHz, it is common for radio sources to have breaks in their spectral indexes \citep{Spectral_index_break} and emission from dust can start to become important.  The principal reason for uncertainty in the effects of compact sources on the $Y$-$M$ relationship is due to the limited number of direct observations at tSZ frequencies.

In this paper, we present results from a large survey for compact sources in tSZ-selected clusters.  Our observations and data reduction are in Sections~\ref{sec:data} and \ref{sec:reduction}.  Section~\ref{sec:counterparts} describes our search for counterparts to the sources we found at other wavelengths and in Section~\ref{sec:comp} we compare our results with the Websky model \citep{Stein2020} for source distributions and the smaller cluster sample in D2021. How the $Y$-$M$ relationship for ACT DR5 is affected by compact sources is covered in Section~\ref{sec:YM} and our conclusions can be found in Section~\ref{sec:conclusions}. We report all errors at a $1\sigma$ / 68.5\% confidence level.

\section{Observations} \label{sec:data}
We surveyed all clusters in the ACT DR5 sample above a declination of $-6^\circ$ and within Right Ascensions between $8^h40^m$--$13^h00^m$ and $2^h00^m$--$3^h00^m$ at $9''$ angular resolution in the continuum at 90~GHz. 
These targets have good visibility from the GBT at night during the high-frequency observing season. To aid in the classification of sources, there is good coverage from other surveys at higher and lower frequencies including the Wide-field Infrared Survey Explorer \citep[WISE;][]{WISE_pntsrc}, 
the Faint Images of the Radio Sky at Twenty-cm (FIRST) point source catalog \citep{FIRST}, and the VLA Sky Survey \citep[VLASS;][]{VLASS}. 

Within the survey footprint there are 300 ACT-identified clusters.  No other cuts were applied to the sample, so our survey accurately reflects the population of ACT DR5 tSZ-selected galaxy clusters.  One notable aspect of the ACT DR5 cluster sample is that it excludes any targets on sight lines to bright point sources ($\geq 10 \, {\rm mJy}$  at 150 GHz).  We refer to this survey as the M2-ACT cluster survey.  

It is important that the survey be made at angular resolutions better than $\sim$15$''$ so that sources can be separated from the larger core size of a galaxy cluster.  Sources up to $4'$ from cluster centers can still have an effect on the measured tSZ (see Figure~\ref{fig:flux2yc}), consequently, an ideal survey to quantify the effects of compact sources would measure a large number of clusters out to this radius, significantly larger than the field-of-view of ALMA. MUSTANG2 (M2) \citep{MUSTANG2}, a camera on the GBT has a $4.'2$ field-of-view, a resolution of $9''$ and can reach noise levels below 60~$\mu$Jy/beam in an hour. MUSTANG2 also has a bandpass well matched to CMB experiments such as ACT thus simplifying data analysis and making it an ideal instrument for this survey.

The M2-ACT cluster survey target depth was a compromise between the desire to detect faint sources that, while common, only affect the measured mass of a cluster by a few percent and observing enough clusters in our allotted telescope time to gain reliable statistics on less common brighter sources that affect the measured mass by far more. To estimate the expected number of sources in the M2-ACT cluster survey, one can scale from the deeper observations of 30 clusters in D2021 which found nine sources brighter than 1~mJy. If the two cluster samples have similar source populations, $90\pm38$ sources of this strength can expected in 300 clusters.  The relatively large errors come from the small D2021 sample size and in this calculation a normal distribution is assumed.  From Equation~\ref{equ:dy0}, a 1~mJy source with a line-of-sight location in the center of a cluster would give a 12\% change in measured Compton-$y$ for a typical ($y_c=7.2\times 10^{-5}$) ACT DR5 cluster.  Consequentially, our target noise was set to be at 200~$\mu$Jy to enable $5\sigma$ detections of 1~mJy sources in the center of each ACT DR5 cluster.  For MUSTANG2, this depth can be achieved with just 5 minutes per cluster. Deeper observations would mean that significantly fewer clusters could be included in the survey.  A wider, shallower survey was not possible due to the limits of the GBT slew rate and the desire to observe each region of the cluster multiple times with independent detectors.   A scan pattern was selected that, in combination with the MUSTANG2 camera's field-of-view, would provide useful coverage out to $\sim$4$'$ in radius for each target.  This comfortably encompasses the region within which discrete sources can affect the ACT DR5 cluster selection.

\begin{table}
\caption{A partial list of ACT DR5 clusters with successful observations and the depth of each map. Compton-$y$ values were taken to be the $y_c$ values in the DR5 catalog. The full table can be found as supplementary material and individual maps can be found on \dataset[dataverse]{https://doi.org/10.7910/DVN/FQNKYX}. 
}\label{tab:clusters}
\hspace{-1.3cm}\resizebox{1.2\linewidth}{!}{%
\begin{tabular}{ccccc}
   ACT DR5 Name & Map Noise & M500c & Compton-$y$ & Redshift \\
        &  (mJy/beam)     & ($\times10^{14}M_\odot$)  & ($\times 10^{-4}$) & \\\hline
ACT-CLJ0200.3$+$0019 & 0.13 & 1.74 & 0.69 & 0.70 \\
ACT-CLJ0201.6$-$0211 & 0.17 & 2.85 & 0.63 & 0.20 \\
ACT-CLJ0201.6$-$0503 & 0.14 & 1.68 & 0.53 & 0.81 \\
ACT-CLJ0203.0$-$0042 & 0.16 & 1.95 & 0.46 & 0.44 \\
\ldots \\
\end{tabular}%
}
\end{table}

\section{Data reduction} \label{sec:reduction}
The standard MUSTANG2 data reduction pipeline \citep[MIDAS; see][for details]{Romero2020} was used to produce maps with 2$''$ pixels for each of our targets. The simulations described below showed that moderately aggressive high-pass filtering of the raw MUSTANG2 time streams at 0.05~Hz produced the best recovery of compact sources. Extra care had to be taken when flagging detector glitches to avoid false positives.  Inclement weather resulted in higher than anticipated noise in some clusters but 243 clusters had noise levels better than our 0.2~mJy RMS target (Table~\ref{tab:clusters}).  For automatic detection of compact sources, a matched filter to the MUSTANG2 beam would be ideal.  However, as the noise properties in the maps vary (both between maps and as a function of radius), such a filter is challenging to calculate.  Instead, we used a difference-of-Gaussians (DoG) filter 
which approximates a matched filter by convolving with two different width Gaussians ($\sigma_1=3.''6$ and $\sigma_2=13.''6$) and taking the difference. These widths were found empirically to produce the strongest detections of point sources injected into our maps. As well as a regular map, a noise map was made by flipping the sign of half of the data from each cluster.  Using the noise map normalized by the integration time at each point, it was possible to make signal-to-noise ratio (SNR) maps for each cluster. 
The signal from sources will always be positive, and for our target survey depth at MUSTANG2's resolution, the negative signal from the galaxy clusters is negligible. This allows sources to be found by looking for positive peaks in the SNR maps.   Once a source location is established, a fit for its amplitude and width was carried out in the original signal map using a 2D Gaussian.  By using the original signal map for this fit, there was no need to account for the effects of the DoG filter.

\begin{figure}
    \centering
    \includegraphics[width=\linewidth]{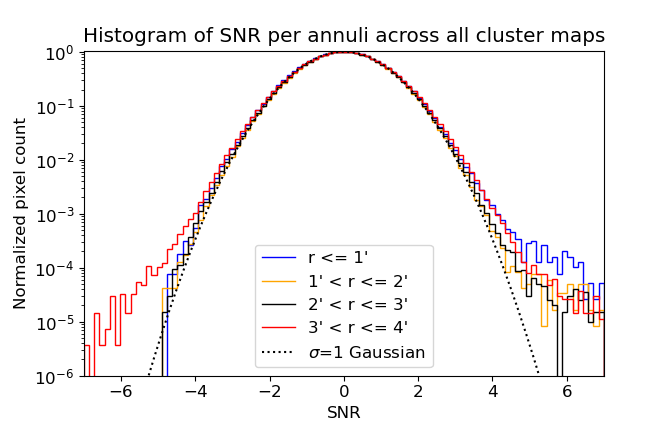}
    \caption{Histograms of the signal-to-noise ratios of  pixels in the MUSTANG2 maps.  Inside a radius of $3'$ the histograms fall off rapidly and a $5\sigma$ cut should produce less than one false positive in our survey which has $\mathcal{O}10^5$ independent beams.  Outside this radius, there are fewer observations per pixel 
    and the noise statistics have a significant non-Gaussian tail. Consequently, a $7\sigma$ cut was needed to obtain the same false-positive rate. The excess power on the positive side is due to compact sources and is not present if these regions are excluded.}\label{fig:noise}
\end{figure} 

\subsection{Signal-to-Noise cuts.}
With any survey, the signal-to-noise  cut is an important choice.  A value too low will result in false detections. A value higher than the lowest cut needed to avoid false detections will result in valid sources being missed. The statistical properties of the pixels in the SNR maps are a good fit to a Gaussian with $\sigma=1$. The M2-ACT survey has approximately $10^5$ independent beams so, if this fit were perfect, a $5\sigma$ cut should give less than one false detection in the entire survey. Initial searches for sources used  this uniform $5\sigma$ cut. However, outside of a radius of $3'$ the number of sources found increased rapidly (even though these areas of each map had higher noise). Most of these sources had a peak SNR below $7\sigma$. Closer examination showed that, due to the lower number of observations in a few pixels, the calculated noise in these areas of the maps has a long non-Gaussian tail (Figure~\ref{fig:noise}). From the negative side of the distribution in SNR values it can be seen that, between 3$'$ and 4$'$, a $7\sigma$ threshold is sufficient for the expected number of false detections to be less than one.  Moreover, sources at $r>3'$ have five times less impact on the measured mass of the cluster (Figure~\ref{fig:flux2yc}) so accepting a slightly higher threshold at $r>3'$ of $7\sigma$ is still consistent with our goal of of characterizing the population all sources that could significantly affect the measured Compton-$y$ of a typical cluster. 

\subsection{Sources found}
An end-to-end test of our pipeline was carried out. Fake sources were placed in noiseless maps at random locations within $4'$ of the cluster centers.  These maps were sampled using the telescope pointing information and the results added to the raw time streams. The fake sources had random amplitudes between 10 and 30~mJy.  Using the same processing steps described above, we were able to recover the amplitudes and locations of all sources without bias, thus confirming the fidelity of our mapmaking   and source finding in the high-SNR limit. To ensure our detection threshold was not too low, our pipeline was run looking for peaks with amplitudes below $-5\sigma$ ($-7\sigma$ in the outer portions of the maps).  No negative peaks were found, so we can conclude that the vast majority of our candidate sources are real. A further null test was made by searching for sources in the noise maps. For the 243 clusters observed under good conditions, no positive or negative sources were found, confirming the quality of these data. Repeating these tests with values of $\sigma$ 0.2 lower resulted in multiple false detections in the noise maps and at negative values of $\sigma$.  Using values 0.2 higher would have resulted in missing 3 sources.  From this, we conclude that our chosen SNR cuts based on Figure~\ref{fig:noise} are close to optimal. 

\begin{figure}
    \centering   
    \includegraphics[width=\linewidth]{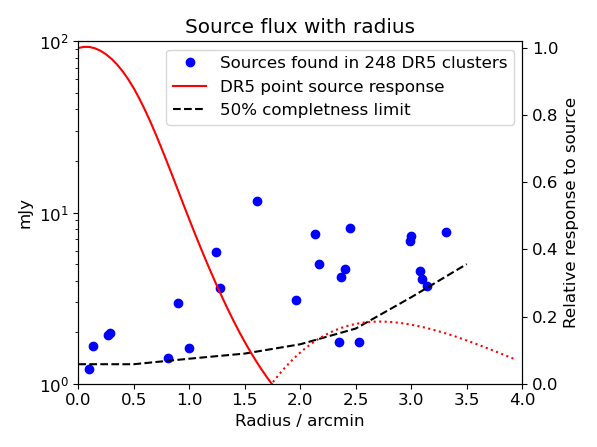}
    \caption{The fluxes and distances from the cluster centers of sources found in our survey of ACT DR5 clusters.  Shown in red is the response of the ACT DR5 cluster pipeline to a compact source.  Solid parts of this line are regions where a source reduces the measured Compton-$y$ while in the dashed region a source increases the measured value.  The black dashed line represents the 50\% completeness limit of our survey, below which our constraining power on the statistical properties of sources becomes unreliable.  As might be expected with a tSZE-selected sample, there are no bright sources in the center, but even so there is a deficit of sources below 10~mJy in the central $1'$.  }\label{fig:sources_found}
\end{figure}

Overall, we found only 24 sources closer to the center of the clusters than $4'$ and with a significance greater than our detection thresholds (Table~\ref{tab:sources}, Figure~\ref{fig:sources_found}), a factor of three less than expected by scaling from the ad hoc sample in \citet{Dicker2021}. Some of this discrepancy can be explained by the completeness of our survey at lower flux levels.  Although 1~mJy is over 5 times the map noise, some fraction of 1~mJy sources will coincide with  negative noise fluctuations and fall below our detection thresholds (completeness or Malmquist bias). This effect will also increase the measured flux of a population of sources near a survey limit, requiring that the measured fluxes be deboosted. To better characterize the completeness and bias, further simulations were carried out. A large number of fake sources with fluxes between 0.5 and 10~mJy were placed at random locations in the final MIDAS maps.  The fraction of these that were detected gave the completeness of our survey, while the fitted amplitudes gave the bias in the measured fluxes. Both of these are functions of radius from the cluster center and flux.   However, even when these effects are taken into account, there are still fewer sources found in the M2-ACT survey than expected.  This is discussed more in Section~\ref{sec:comp}.

\newcommand{\figfourcaption}{Signal to noise plots of our sources and their corresponding spectral energy densities (SEDs).  The MUSTANG2 source is shown as the red circle along with its detected SNR value.  Sources from other point source catalogs are shown as the same symbols used on the SED plots. Although error bars are plotted on the SED plot in many cases they are not visible.  When a point source catalog did not contain a counterpart the approximate detection threshold is shown as a triangle with a one sided error bar. The fitted radio spectral index between MUSTANG2 and FIRST (or the average spectral index from D2021 when no FIRST counterpart was found) is shown as the blue dotted line. The green dashed line shows a typical 40K dust spectrum that goes through the MUSTANG2 point.  The SPIRE detections or upper limits are mostly below the dust spectrum curves, indicating that MUSTANG2 sources are mostly dominated by radio emission.  The WISE and 2MASSS data indicate the presence of hot gas in some sources, but this will have a negligible contribution at 90~GHz.}

\begin{figure}
    \centering
    \includegraphics[width=0.95\linewidth]{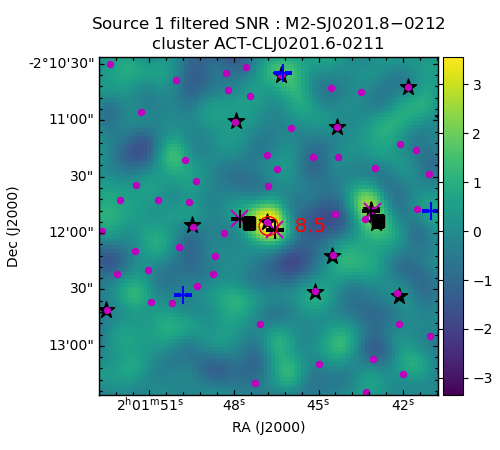}
    \includegraphics[width=0.95\linewidth]{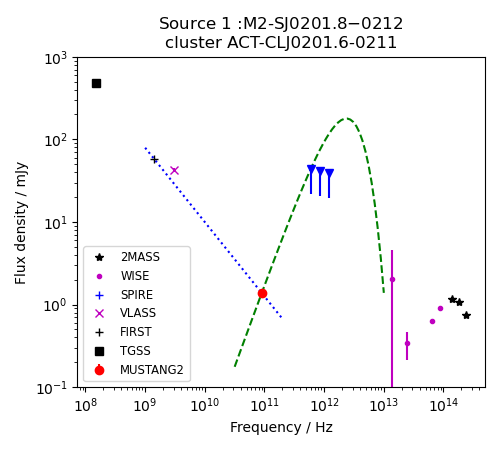}
    \caption{An SNR plot of one of our sources (top) and the corresponding SED plot (bottom).  On the SNR plot, the MUSTANG2 source is shown as the red circle along with its detected SNR value.  Sources from other point-source catalogs are shown as the same symbols used on the SED plots. Although error bars are plotted on the SED plot, in many cases, they are not visible.  When a point-source catalog did not contain a counterpart, the approximate detection threshold is shown as a triangle with a one-sided error bar. The fitted radio spectral index between MUSTANG2 and FIRST (or the median spectral index from D2021 when no FIRST counterpart was found) is shown as the blue dotted line. The green dashed line shows a typical 40~K dust spectrum that goes through the MUSTANG2 point.  The SPIRE detections or upper limits are mostly below the dust spectrum curves, indicating that MUSTANG2 sources are mostly dominated by radio emission.   The WISE and 2MASS data indicate the presence of hot gas in some sources, but this will have a negligible contribution at 90~GHz.  Similar plots for all 24 sources found are 
    shown at the end of the pdf. 
    }\label{fig:examplemap}
\end{figure}

\section{Source Counterparts}\label{sec:counterparts}
The list of candidate point sources was matched to 
point-source catalogs.  If the angular resolution of the catalog was greater than MUSTANG2, a matching radius of 9$''$ was used.   If the catalog was from a lower-resolution survey, the resolution of the survey was used.   Spectral energy density (SED) plots for each source were made --- see Figure~\ref{fig:examplemap} for an example.  

\subsection{Radio counterparts}
Searches for radio counterparts were carried out using catalogs made from the 150~MHz TGSS \citep{TGSS}, the 1.4~GHz FIRST survey \citep{FIRST}, and the 2--4~GHz VLASS survey \citep{VLASS}.  These surveys have detection limits of approximately 24.5~mJy, 1~mJy, and 0.5~mJy, respectively. Using a search radius of 9$''$, matches were found for 75\% of our detected sources.  In two cases, two counterparts were found within our search radius.  As these were within one MUSTANG2 beam, the sum of the total integrated flux was used when plotting spectral energy density (SED) plots and calculating spectral indexes for use in Equation~\ref{equ:dy0}.  

\begin{figure}
\centering
\includegraphics[width=\linewidth]{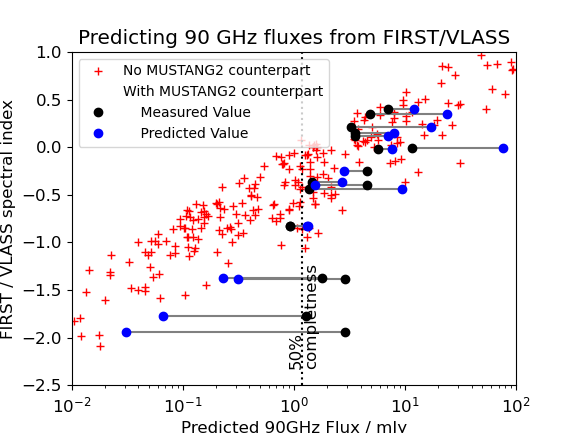}
    \caption{The 90~GHz fluxes predicted using all pairs of sources from the FIRST and VLASS catalogs within our clusters.  Red crosses represent those not detected by MUSTANG2 while blue dots represent detections.  For all detections, the measured value is shown as a black point and is often higher or lower by an order of magnitude. This failure to predict the correct flux, especially the large number of predicted fluxes above our detection threshold, demonstrates the need for the higher-frequency measurements in this paper.
    }\label{fig:FIRST_VLASS}
\end{figure}

An effort was made to see if the FIRST and VLASS catalogs could be used to predict which sources could have fluxes at 90~GHz large enough to affect Compton-$y$ measurements.  For all sources that appeared in both catalogs and within $4'$ of the center of one of the clusters in our sample, the flux at 90~GHz was predicted assuming a constant spectral index between FIRST and 90~GHz.  As shown in Figure~\ref{fig:FIRST_VLASS},  there is little difference in the distributions of  either the predicted 90~GHz flux or the spectral index between those sources that are counterparts to MUSTANG2 and those that are not seen.  Of the sources that are seen by MUSTANG2, some counterparts predict lower 90~GHz fluxes (possibly due to dust), while others predict higher fluxes indicating a steepening of the spectral index that is often seen in radio sources at higher frequencies \citep{Spectral_index_break}. Any source variability could also cause discrepancies between measured and predicted 90~GHz fluxes. However, there are a large number of sources with predicted fluxes well above the M2-ACT detection limit that are not seen -- far larger than could be explained by source variability (which would cause a random spread).  This is best explained by most sources having a steepening of the spectral index toward high frequencies. 

Possible changes in spectral index at 1.4~GHz and at 90~GHz and the fact that not every source had a counterpart strongly argue for high-resolution follow-up at other tSZ frequencies such as 150~GHz.  As complete surveys of every cluster are not needed, such observations could be made quickly.

\subsection{Higher-frequency counterparts}
Submillimeter and infrared continuum emission from galaxies is normally well described by a so-called greybody, with $25~{\rm K} \lesssim T \lesssim 50{\rm ~K}$ and $\beta \sim 1.5$.  Available point-source catalogs at these frequencies include SPIRE \citep[600--1200~GHz:][]{SPIRE}, WISE \citep[12--89~THz:][]{WISE_pntsrc}, and the Two Micron All Sky Survey \citep[2MASS, 138--240~THz:][]{2MASS_catalog}. 
\begin{rotatetable*}
\begin{deluxetable*}{llcccccl}
\tablefontsize{\footnotesize}
    \tablecaption{Sources found by our survey.  None of the sources were strong enough to enable fitting for extended emission, so only peak fluxes are quoted. A machine readable version of this table that includes the fluxes of counterparts can be found as supplementary material. 
    }\label{tab:sources}
\tablehead{\colhead{} & \colhead{ACT} & \colhead{RA} & \colhead{Dec} & \colhead{Flux} & 
\colhead{Radius}  & \colhead{Spectral} & \colhead{} \\
\colhead{source ID} & \colhead{cluster} & \colhead{J2000} & \colhead{J2000} & \colhead{(mJy)} & 
\colhead{from center}  & \colhead{index} & \colhead{Counterparts}
}
\startdata
M2-SJ0201.8$-$0212 & ACT-CLJ0201.6-0211 & 02h01m46.8s & $-$02$^\circ$11$'$56$''$ & 1.38$\pm$0.14 & 1.00$'$ & -0.90  & TGSS,FIRST,VLASS,WISE,2MASS\\
M2-SJ0201.7$-$0212 & ACT-CLJ0201.6-0211 & 02h01m43.3s & $-$02$^\circ$11$'$45$''$ & 1.46$\pm$0.16 & 0.14$'$ & -0.52  & FIRST,VLASS,WISE,2MASS\\
M2-SJ0204.5$+$0320 & ACT-CLJ0204.5+0321 & 02h04m29.3s & $+$03$^\circ$19$'$33$''$ & 8.00$\pm$0.38 & 2.44$'$ & 0.27  & FIRST,VLASS,WISE,2MASS\\
M2-SJ0208.5$-$0259 & ACT-CLJ0208.3-0255 & 02h08m27.7s & $-$02$^\circ$59$'$10$''$ & 7.48$\pm$0.59 & 3.31$'$ & -0.46  & \\
M2-SJ0220.8$-$0333 & ACT-CLJ0220.9-0332 & 02h20m49.5s & $-$03$^\circ$33$'$12$''$ & 2.93$\pm$0.27 & 1.96$'$ & -0.85  & TGSS,FIRST,VLASS,WISE\\
M2-SJ0223.7$+$0019 & ACT-CLJ0223.6+0020 & 02h23m39.3s & $+$00$^\circ$18$'$51$''$ & 7.44$\pm$1.05 & 2.13$'$ & -0.46  & WISE\\
M2-SJ0231.7$-$0453 & ACT-CLJ0231.7-0453 & 02h31m41.3s & $-$04$^\circ$52$'$43$''$ & 2.86$\pm$0.27 & 0.89$'$ & -0.46  & \\
M2-SJ0233.3$+$0446 & ACT-CLJ0233.2+0448 & 02h33m16.5s & $+$04$^\circ$45$'$46$''$ & 6.63$\pm$0.68 & 2.99$'$ & -0.46  & WISE \\
M2-SJ0248.1$-$0332 & ACT-CLJ0248.0-0331 & 02h48m03.3s & $-$03$^\circ$31$'$40$''$ & 0.93$\pm$0.07 & 0.10$'$ & -0.91  & TGSS,FIRST,VLASS,WISE,2MASS\\
M2-SJ0248.4$-$0334 & ACT-CLJ0248.3-0337 & 02h48m21.9s & $-$03$^\circ$34$'$21$''$ & 3.73$\pm$0.24 & 3.10$'$ & -0.20  & TGSS,FIRST,VLASS,WISE,2MASS\\
M2-SJ0249.6$+$0212 & ACT-CLJ0249.6+0210 & 02h49m36.9s & $+$02$^\circ$12$'$15$''$ & 1.36$\pm$0.15 & 2.35$'$ & -0.46  & \\
M2-SJ0839.7$-$0153 & ACT-CLJ0839.5-0150 & 08h39m40.3s & $-$01$^\circ$52$'$33$''$ & 7.14$\pm$0.34 & 3.00$'$ & 0.28  & FIRST,VLASS,WISE\\
M2-SJ0839.7$-$0150 & ACT-CLJ0839.5-0150 & 08h39m43.4s & $-$01$^\circ$50$'$19$''$ & 4.21$\pm$0.38 & 3.08$'$ & 0.37  & FIRST,WISE,2MASS\\
M2-SJ0901.4$+$0259 & ACT-CLJ0901.5+0301 & 09h01m26.1s & $+$02$^\circ$59$'$26$''$ & 4.55$\pm$0.37 & 2.40$'$ & -0.14  & FIRST,VLASS,WISE,2MASS\\
M2-SJ0905.7$+$0434 & ACT-CLJ0905.6+0434 & 09h05m43.9s & $+$04$^\circ$33$'$42$''$ & 5.80$\pm$0.23 & 1.24$'$ & -0.09  & TGSS,FIRST,VLASS,WISE,2MASS\\
M2-SJ0918.8$+$0213 & ACT-CLJ0918.7+0211 & 09h18m48.8s & $+$02$^\circ$13$'$26$''$ & 4.87$\pm$0.21 & 2.17$'$ & -0.03  & FIRST,VLASS,WISE,2MASS\\
M2-SJ1118.0$-$0211 & ACT-CLJ1117.9-0211 & 11h17m59.7s & $-$02$^\circ$11$'$07$''$ & 1.74$\pm$0.17 & 0.27$'$ & -0.09  & \\
M2-SJ1148.2$-$0236 & ACT-CLJ1148.0-0234 & 11h48m14.5s & $-$02$^\circ$36$'$03$''$ & 1.30$\pm$0.18 & 2.53$'$ & -1.10  & TGSS,FIRST,VLASS,WISE\\
M2-SJ1152.4$+$0029 & ACT-CLJ1152.2+0031 & 11h52m22.5s & $+$00$^\circ$28$'$57$''$ & 3.27$\pm$0.29 & 3.14$'$ & -0.19  & TGSS,FIRST,VLASS\\
M2-SJ1204.0$+$0127 & ACT-CLJ1203.9+0126 & 12h04m02.2s & $+$01$^\circ$26$'$42$''$ & 3.54$\pm$0.20 & 1.28$'$ & -0.04  & FIRST,VLASS,WISE,2MASS\\
M2-SJ1210.3$+$0224 & ACT-CLJ1210.2+0223 & 12h10m18.0s & $+$02$^\circ$23$'$40$''$ & 1.81$\pm$0.29 & 0.28$'$ & -0.88  & TGSS,FIRST,VLASS,WISE,2MASS\\
M2-SJ1220.3$-$0241 & ACT-CLJ1220.4-0240 & 12h20m18.5s & $-$02$^\circ$40$'$58$''$ & 11.67$\pm$0.22 & 1.61$'$ & -0.60  & TGSS,FIRST,VLASS,WISE,2MASS\\
M2-SJ1224.5$+$0211 & ACT-CLJ1224.3+0211 & 12h24m27.9s & $+$02$^\circ$11$'$03$''$ & 4.04$\pm$0.34 & 2.37$'$ & -1.20  &  VLASS\\
M2-SJ1401.0$+$0252 & ACT-CLJ1401.0+0252 & 14h01m00.8s & $+$02$^\circ$51$'$41$''$ & 1.15$\pm$0.16 & 0.81$'$ & -0.08  & FIRST,WISE
\enddata
    \end{deluxetable*}
\end{rotatetable*}
SED plots in the extended version of Figure~\ref{fig:examplemap} show that the majority of the MUSTANG2 detections have WISE counterparts. However, searches carried out at random locations on each map showed that over 25\% of these matches could be line-of-sight coincidences.  The 2MASS counterparts seen in 50\% of the MUSTANG2 sources are more reliable, and their amplitudes point toward a hotter dust component that would have an insignificant amount of emission at 90~GHz.  Not all the clusters in our ACT sample had data from SPIRE, but for those that did have coverage, few SPIRE sources were counterparts to MUSTANG2.  However, it was possible to place upper limits on the 600--1200~GHz emission.  In most cases, these limits were far below what would be expected if the MUSTANG2 sources were dominated by cold dust (Figure~\ref{fig:examplemap}).  This constraint, along with those from 2MASS, implies that the majority of the flux MUSTANG2 sees is radio emission, consistent with the findings from \citet{Dicker2021} and \citealt{OrlowskiScherer2021}. 

\section{Comparisons with Other cluster samples}\label{sec:comp}
\begin{figure}
    \centering
    \includegraphics[width=\linewidth]{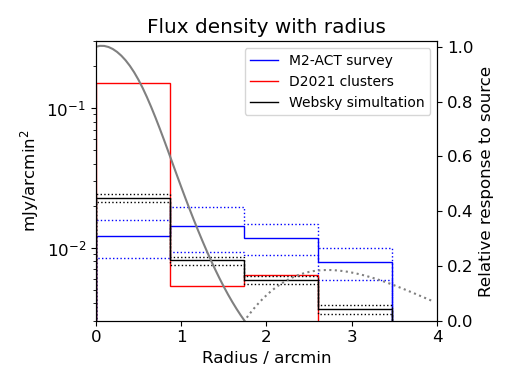}
    \caption{The average 90~GHz flux from point sources as a function of radius for this work (M2-ACT), the D2021 sample, and the Websky simulation. For a fair comparison, the M2-ACT survey completeness function has been applied to the other two samples. Errors have been estimated by bootstrapping techniques and are shown as dashed lines for the Websky and M2-ACT samples.  In the central $52''$, the Websky simulation shows a significant ($>3\sigma$) excess of flux compared to the ACT sample. However, further out, the ACT sample has a 1.3--2$\sigma$ excess of flux from radio sources.  The clear excess of sources in the D2021 sample can be seen in the central bin.  For reference, the response to point sources of the ACT DR5 cluster finder is shown in gray.  }
    \label{fig:flux_w_radius}
\end{figure}

\subsection{Comparisons with D2021.}
The D2021 clusters were all deep observations in order to study the physics of the ICM or to measure the cluster masses.  Many clusters in this nonuniform sample were selected for study because they were known to be mergers or at high redshifts and are atypical. 
 The total fraction of clusters containing sources in the D2021 clusters is higher than that of the M2-ACT sample.  However, the significance of this is highly dependent on assumptions such as the underlying distribution of source fluxes.  Using only the total number of sources found, the hypothesis that the D2021 and the M2-ACT survey samples are drawn from the same parent sample of clusters can be rejected by as little as $1.2\sigma$ to well over $3\sigma$.

It is far more instructive to look at the fluxes and locations of the sources in each sample. The D2021 sample shows a much greater density of sources in the cluster centers (Figure~\ref{fig:flux_w_radius}). This persists even when the completeness limits of this work are taken into account.  Interestingly, this is not the case outside the central 52$''$, where the ACT clusters in this work show more radio emission from sources. Sources outside of $104''$ increase the measured Compton-$y$ and thus increase the likelihood of a cluster being detected by experiments such as ACT. The different distributions of sources within each cluster sample is further evidence that they are not drawn from the same parent cluster sample. Because of the unrepresentative nature and small size of the D2021 sample, it should not be used to correct $Y$-$M$ relationships.

\subsection{Comparisons with Websky}\label{sec:websky}
Websky is a set of simulated observations of the extragalactic microwave sky \citep{Stein2020}. It is used for forecasting for next-generation CMB experiments \citep[e.g.,][]{Hensley2022}, evaluating biases in analysis methods \citep[e.g.,][]{Coulton2023}, and providing theory foregrounds \citep[e.g.,][]{Madhavacheril2023, Qu2023}. The Websky maps come with a corresponding catalog of radio sources \citep{Li2022} and a map of their millimeter emission. This catalog is generated from the halos present in the base Websky observations and whose distribution of intrinsic fluxes and spectral indexes has been calibrated to match observed literature values in the millimeter. One of the primary uses of this radio catalog is to understand biases imparted in tSZ and lensing measurements by radio sources. While the radio fluxes do not bias the lensing reconstruction itself, when using lensing mass calibrations, clusters must be binned in mass to achieve sufficient SNR. This binning is done using the tSZ-inferred mass and hence can be biased by radio source contamination \citep{Madhavacheril2023}. Therefore, confirming that the Websky radio accurately reproduces the observed tSZ contamination is of high priority. 

To this end, we compare the results of the M2-ACT survey to the corresponding Websky simulations. Specifically, we select a $\sim$10,000 cluster subsample of the Websky halos that matches the redshift and mass distributions of the ACT DR5 cluster catalog used in this work. We then identify all Websky radio sources lying within $5'$ radially of the center of the cluster and with flux at $90$\,GHz $\geq0.5$~mJy. This flux cut is low enough so that the effect on the $Y$-$M$ relationship (Section~\ref{sec:YM}) is insensitive to it.  To better compare the samples, the M2-ACT completeness function is then applied.  As can be seen in Figure~\ref{fig:flux_w_radius}, with the completeness applied to the Websky sample, it shows a $3\sigma$ excess of flux compared to the M2-ACT sample within the central $52''$ and as with D2021, outside this radius, there is 1--2$\sigma$ of excess emission from radio sources from the clusters in the M2-ACT sample.

There are two potential explanations for this.  First, the Websky simulation always places radio sources at the centers of the clusters. The only sources further out are those not associated with that cluster.  In practice, many real clusters contain point sources away from their centers. This is especially true at higher redshifts, where sources may still be infalling toward the cluster centers. For this to be the explanation, the Websky data should show an excess of flux in their centers equal to that missing further out.  Although there is an excess of flux in the central $52''$ of the Websky clusters, there is not enough of an excess to explain the excess of sources found between 52$''$ and 3$'$ in the M2-ACT sample.  If the M2-ACT sources are split into a higher- and lower-redshift groups (at $z=0.4$), then it becomes clear that the majority of the sources observed away from the cluster centers are at higher redshifts.  The Websky simulations do not capture this change in source distribution with redshift well, but as the excess of flux in their centers is not equal to the deficit further out, the different distributions of sources within the clusters cannot be the only explanation. 

The second explanation is due to how a source affects the measured Compton-$y$.  For the ACT DR5 pipeline, sources less than 104$''$ from the cluster center (the null in Figure~\ref{fig:flux2yc}) reduce the measured Compton-$y$ so some clusters with strong sources in the center will be missed by tSZ surveys such as ACT.  Sources further out increase the measured Compton-$y$ and could scatter lower-mass clusters into the tSZ-selected cluster sample.  Compared to the underlying cluster population, a tSZ-selected cluster sample can be expected to have an underrepresented  number of clusters with  bright sources in their centers and an overrepresentation of those further out (past 104$''$).  A more quantitative analysis follows.    

\begin{figure}
    \centering
    \includegraphics[width=\linewidth]{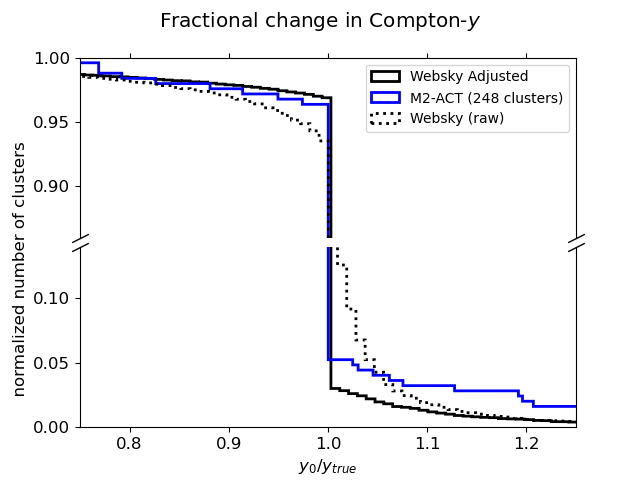}
    \caption{Cumulative plots of the fractional change in measured Compton-$y$ caused by sources calculated using 
    Equation~\ref{equ:dy0}.  The blue curve uses sources in this work and results using sources from the Websky simulation are in black. The dotted black line shows the results using all sources in the Websky sample above 0.5~mJy, while the solid line shows the the same sample taking into account the M2-ACT survey depth. \explain{figure style has been changed to make use of a broken y axis}}\label{fig:dy}
\end{figure}

\section{Implications for the ACT $Y$-$M$ relation}\label{sec:YM}
\explain{this was section heading 5.2.1}
For each cluster in the Websky and ACT samples, $\Delta y_c$ can be calculated using Equation~\ref{equ:dy0} and the locations and fluxes of any sources. As the Websky sample contains many sources below the sensitivity of the ACT sample, we also applied the completeness function described in Section~\ref{sec:reduction} to the Websky sources and recalculated $\Delta y_c$. A comparison of these distributions is shown in Figure~\ref{fig:dy}. Note that this comparison only includes Websky radio sources, and not dusty ones. However, \citealt{OrlowskiScherer2021} found that 
contamination by infrared sources of ACT clusters is negligible as compared to that from radio sources. A similar result was found by \citet{bleem2023} in data from the SPT.

First, we examine the effects of sources in the centers of the clusters that reduce the measured $y_c$ and are responsible for the left side of Figure~\ref{fig:dy}.
Once the Websky sample is adjusted to account for the survey limits of our sample, there is very good agreement between the two when the measured $y_c$ is between 0.8 and 1.0 times the true value. Although this may seem like a contradiction to the excess flux shown in the Websky sample between $0''$ and $52''$ in Figure~\ref{fig:flux_w_radius}, once the response to a point source and the excess flux in the M2-ACT sample between $52''$ and $104''$ is taken into account, the results are consistent. The Websky sample also predicts that 1.5\% of the clusters should have a measured $y_c$ less than 0.75 times its true value. No ACT clusters with sources strong and close enough to their centers to produce this large of an effect were observed. If the Websky survey distribution of such strong central sources were correct, this would imply that tSZ surveys will miss 1.5\% of clusters.  Without more data from a non-tSZ selected cluster sample it is not possible to confirm this number.  For this reason, follow-up observations of an X-ray-selected cluster sample are planned. 

Without the flux cut, the Websky sample predicts that 5\% of clusters should have sources that reduce $y_c$ by more than 5\%.  The difference between Websky predictions of $\Delta y_c$ with and without the flux cut represents the contribution of sources with fluxes just under 1~mJy.  Although the ACT sample measured in this paper cannot place good limits on fainter sources, it shows a good match to brighter sources from Websky in the center of the clusters. Consequently, it is likely that 5\% of clusters have measured $y_c$ values 5\% or more lower than their true value.   

At larger distances from the cluster center, sources increase the measured $y_c$ from its true value.  In Figure~\ref{fig:dy} this shows up on the right. There is disagreement between the Websky prediction and our sample.  Our sample shows that 3.3\% of clusters have measured $y_c$ 10\% or more greater than the true value and 2\% have values 20\% greater or more, while Websky gives numbers of 1.5\% and 0.7\%, respectively.  Some clusters in our sample had values of $y_c$ 50\% high. 

\section{Conclusions}\label{sec:conclusions}
Our survey of 243 ACT clusters detected 24 sources. Even allowing for different survey depths, this is significantly lower than expected when extrapolating from D2021. Most (75\%) of these sources had low-frequency radio counterparts, but deep, high-resolution follow-up at other tSZ bands would help confirm detections and spectral indexes. Within the central $104''$ of the clusters, there is a reasonable agreement on the effect on $y_c$ with Websky for sources in the 1--5~mJy range (which reduce the measured $y_c$ by up to 80\%).  Missing from our survey are clusters with strong sources in their centers, likely because such sources would lower $y_c$ enough to exclude clusters from our tSZ-selected cluster sample.  A complimentary survey of X-ray-selected clusters is planned in order to place limits on the number of bright 90~GHz sources in the centers of clusters.  

At larger radii from the cluster centers, our survey found an excess of sources brighter than $\sim$1~mJy compared to Websky.  The planned X-ray sample will distinguish if this is a limitation in the Websky model or a selection effect of the tSZ sample.  Regardless of the cause, the Websky model does not reproduce the same positive changes in $y_c$. If a deeper MUSTANG2 survey had been possible, it is likely the discrepancy with Websky would have become larger.  Using a simple extrapolation to fainter sources based on the relative numbers of fainter Websky sources, we calculate that  over 3\% of clusters would have a 20\% or more increase in measured $y_c$.  In a stacking analysis of our results, sources in the outer regions increase the average mass bin by 1\% (compared to 0.3\% for the Websky prediction). While small, this value could easily increase to 2\% once fainter sources are included, a level that if unaccounted for, is comparable to the desired accuracy of future tSZ surveys.  Once our X-ray survey is complete, comparison with other source models such as Sehgal \citep{Sehgal} or Agora \citep{Agora} simulations would be interesting.

When working with individual cluster masses from tSZ surveys, there is a chance of very significant ($>$30\%) errors in the individual masses due to radio sources.  Extrapolations from low-frequency radio surveys do not predict this well.  With observations often taken decades apart, source variability may play a role here, but by far the biggest effect is the huge range in spectral indexes of the sources.  Confirmation of masses by other methods such as weak lensing or searches for sources with better than 1~mJy sensitivity at frequencies around tSZ wavelengths would be needed for science results that depend on the mass of an individual cluster.

In summary, the Websky model for point sources is sufficient for today's cluster surveys, but in the future, such models may need to include offsets of the sources from the tSZ centers.  Predicting the distributions of sources from low-frequency data is not possible.  Instead, high-resolution observations in the tSZ frequency bands are needed to confirm models for point-source distributions in clusters.  These observations should include non-tSZ cluster samples and measurements at multiple frequencies in order to better classify source properties.

\section*{acknowledgments}
Construction and early operations of MUSTANG2 were supported by NSF award No. 1615604 and the Mt.\ Cuba Astronomical Foundation. The GBT data were acquired under the project IDs AGBT21B\_298, and AGBT22B\_242.  The Green Bank Observatory is a facility of the National Science Foundation operated under cooperative agreement by Associated Universities, Inc. This publication makes use of data products from the Wide-field Infrared Survey Explorer, which is a joint project of the University of California, Los Angeles, and the Jet Propulsion Laboratory/California Institute of Technology, funded by the National Aeronautics and Space Administration.
S.R.D. is supported by NSF grant No. 2307546. M.M. acknowledges support from NSF grants AST-2307727 and  AST-2153201 and NASA grant 21-ATP21-0145. 
M.H. acknowledges financial support from the National Research Foundation of South Africa.
L.D.M. acknowledges support from the French government, through the UCA\textsuperscript{J.E.D.I.} Investments in the Future project managed by the National Research Agency (ANR) with the reference No. ANR-15-IDEX-01.


%

\vspace{5mm}
\facilities{GBT (MUSTANG2), WISE, Herschel (SPIRE)}


\software{astropy \citep{2013A&A...558A..33A,2018AJ....156..123A}}




\bibliography{m2actSrcs}{}
\bibliographystyle{aasjournal}



\include{Expanded_fig4}


\end{document}

%% file: Expanded_fig4.tex
\setcounter{page}{1}
\renewcommand{\thefigure}{4a}
  \begin{figure*}
    \centering
    \includegraphics[width=0.23\linewidth]{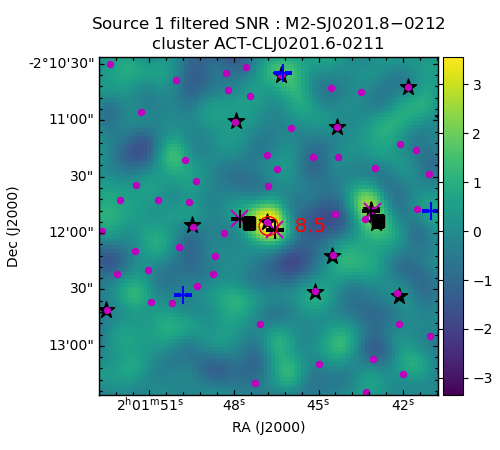}
    \includegraphics[width=0.23\linewidth]{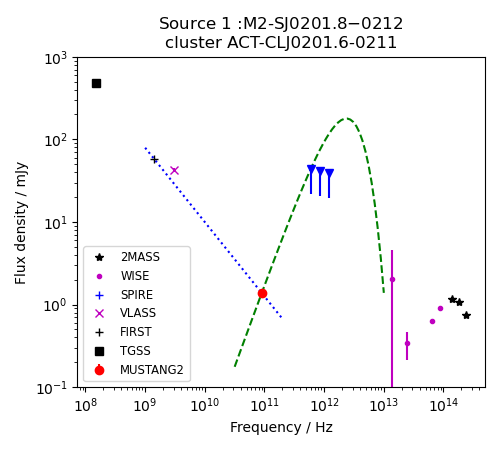}\hspace{0.8cm}
    \includegraphics[width=0.23\linewidth]{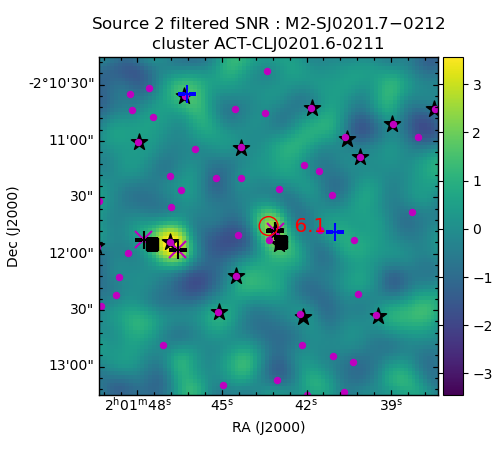}
    \includegraphics[width=0.23\linewidth]{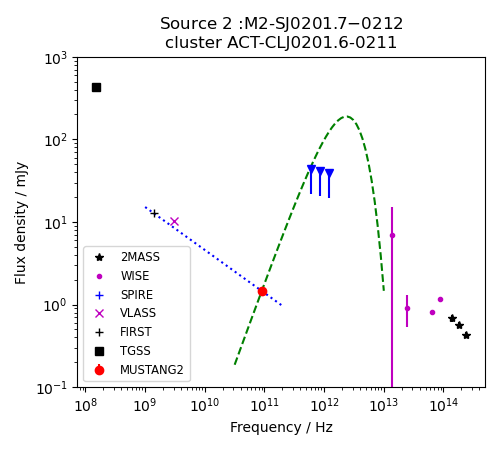}
    
    \includegraphics[width=0.23\linewidth]{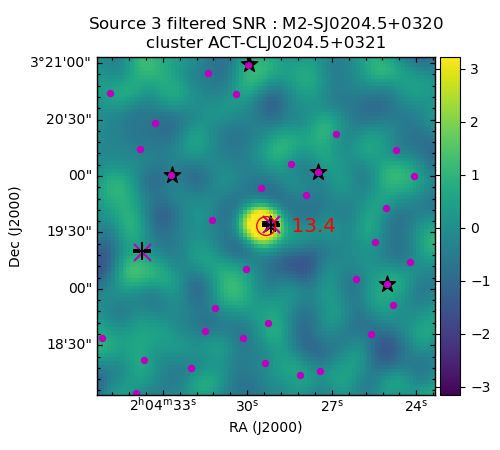}
    \includegraphics[width=0.23\linewidth]{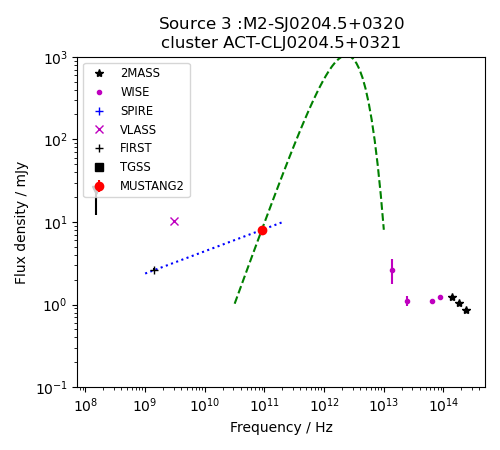}\hspace{0.8cm}
    \includegraphics[width=0.23\linewidth]{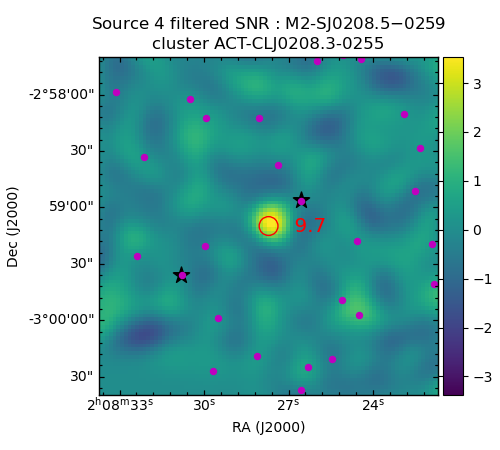}
    \includegraphics[width=0.23\linewidth]{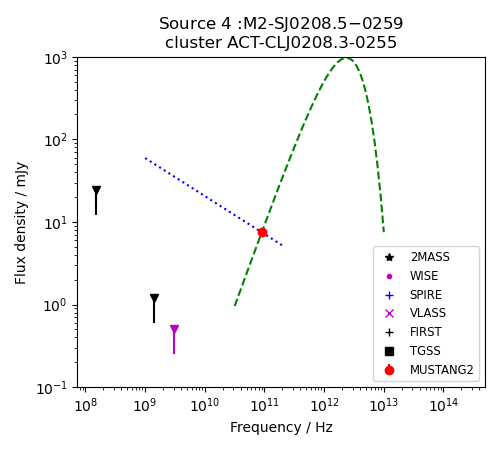}
    
    \includegraphics[width=0.23\linewidth]{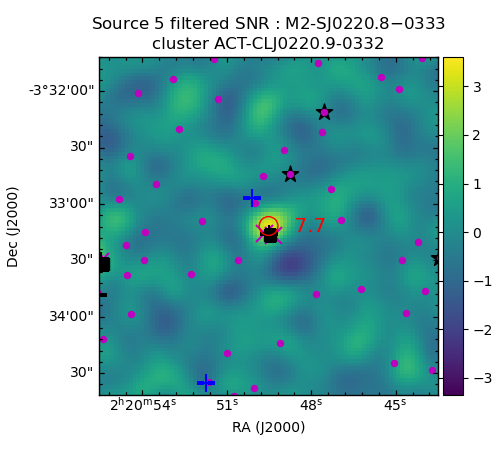}
    \includegraphics[width=0.23\linewidth]{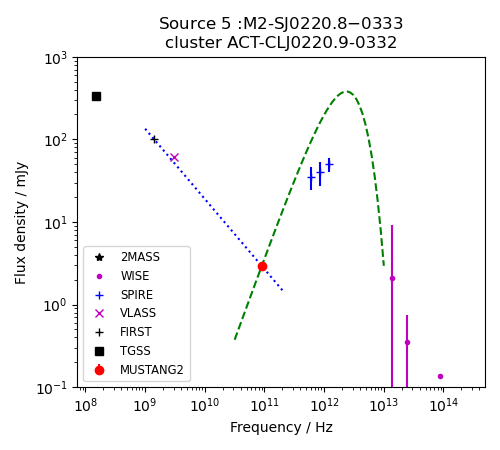}\hspace{0.8cm}
    \includegraphics[width=0.23\linewidth]{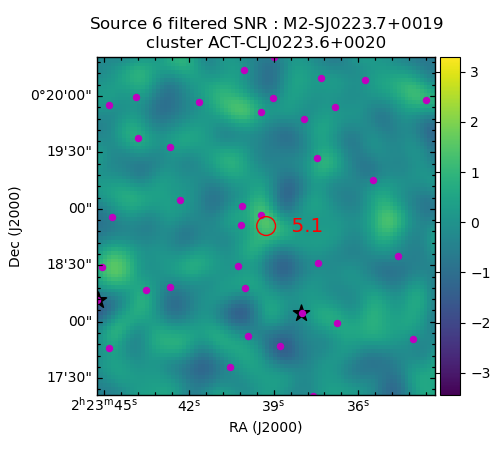}
    \includegraphics[width=0.23\linewidth]{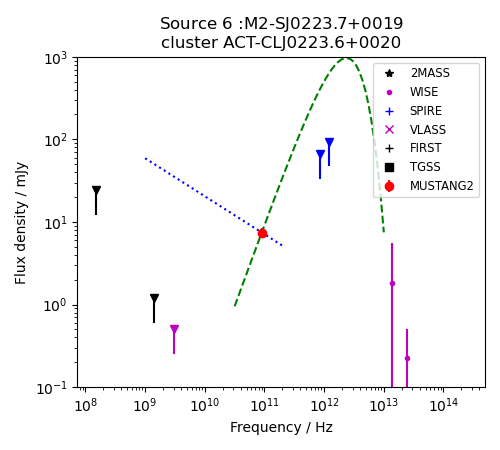}
    
   \includegraphics[width=0.23\linewidth]{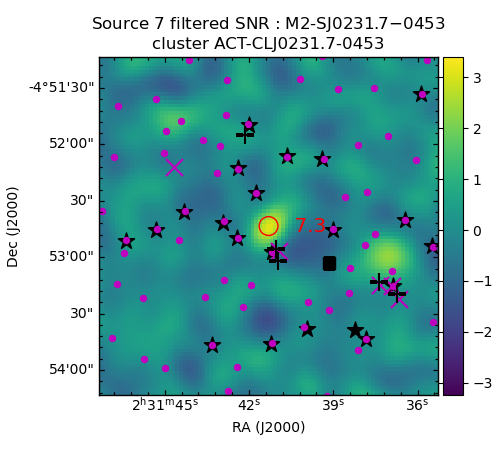}
    \includegraphics[width=0.23\linewidth]{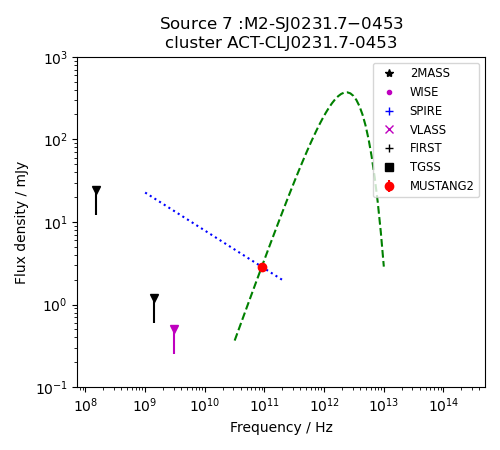}\hspace{0.8cm}
    \includegraphics[width=0.23\linewidth]{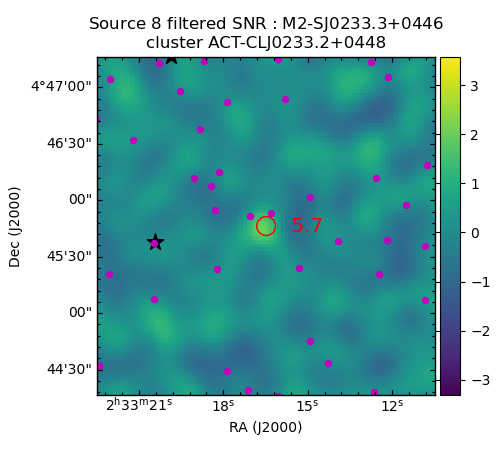}
    \includegraphics[width=0.23\linewidth]{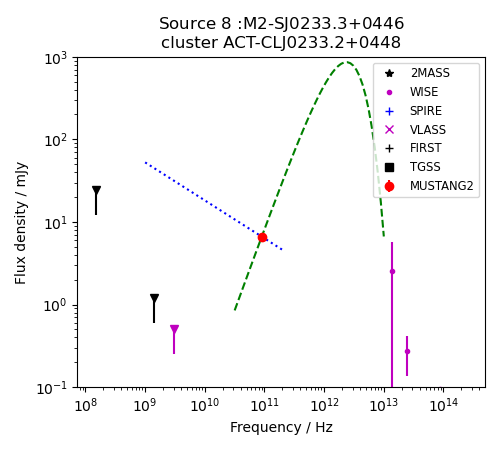}
    
   \includegraphics[width=0.23\linewidth]{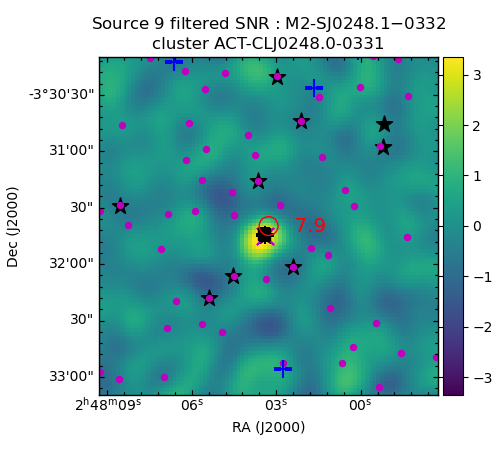}
    \includegraphics[width=0.23\linewidth]{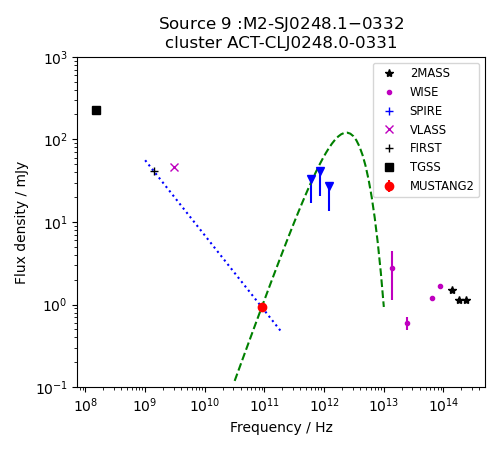}\hspace{0.8cm}
    \includegraphics[width=0.23\linewidth]{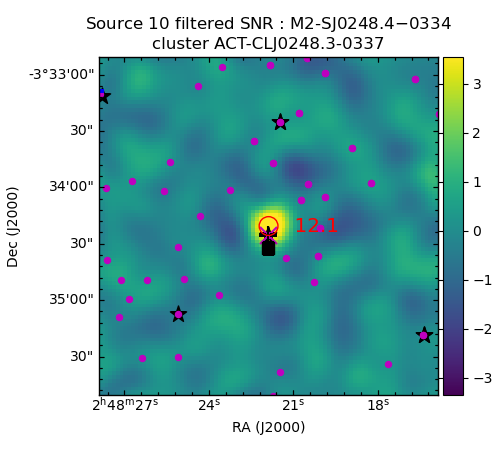}
    \includegraphics[width=0.23\linewidth]{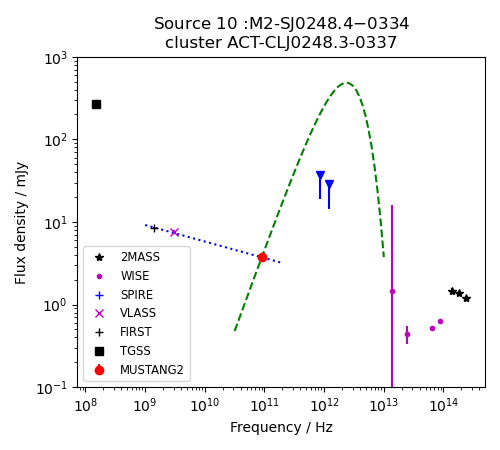}

    \caption{On-line Extra: \figfourcaption \newline
    }
\end{figure*}
\renewcommand{\thefigure}{4b}
\begin{figure*}
    \centering
    \includegraphics[width=0.23\linewidth]{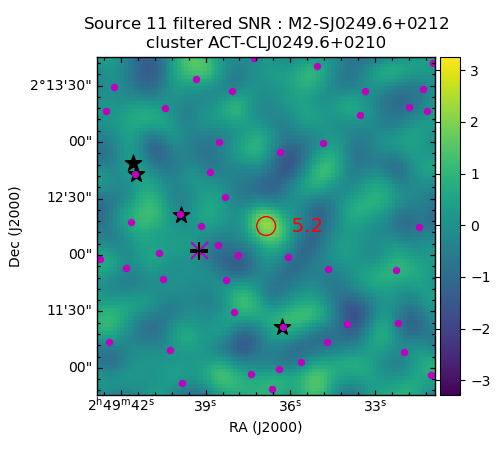}
    \includegraphics[width=0.23\linewidth]{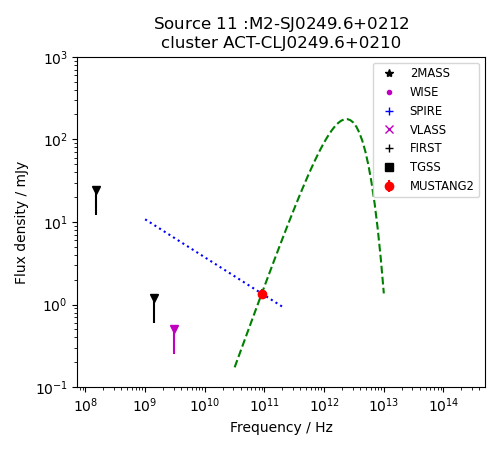}
    \hspace{0.8cm}
    \includegraphics[width=0.23\linewidth]{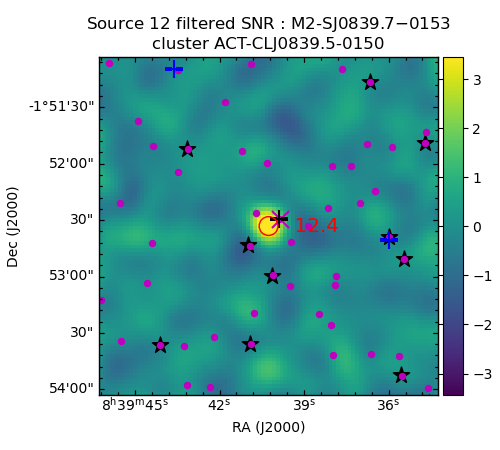}
    \includegraphics[width=0.23\linewidth]{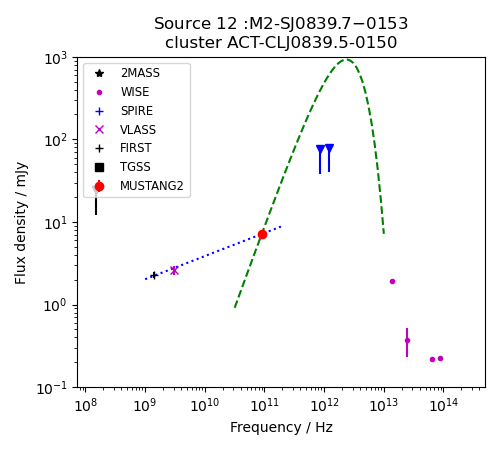}
    
    \includegraphics[width=0.23\linewidth]{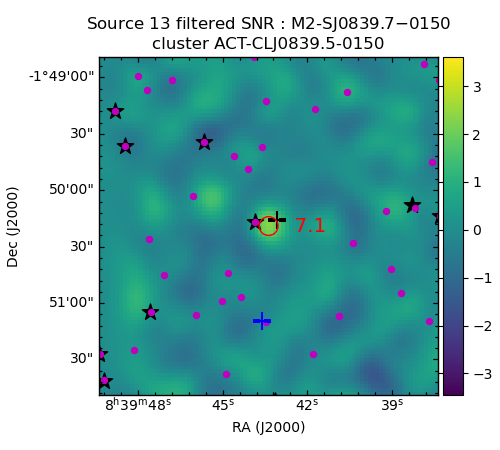}
    \includegraphics[width=0.23\linewidth]{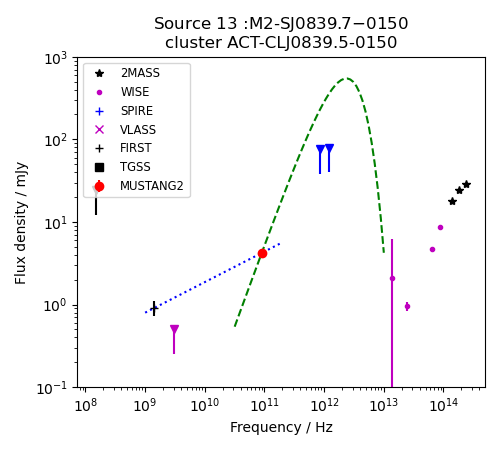}\hspace{0.8cm}
    \includegraphics[width=0.23\linewidth]{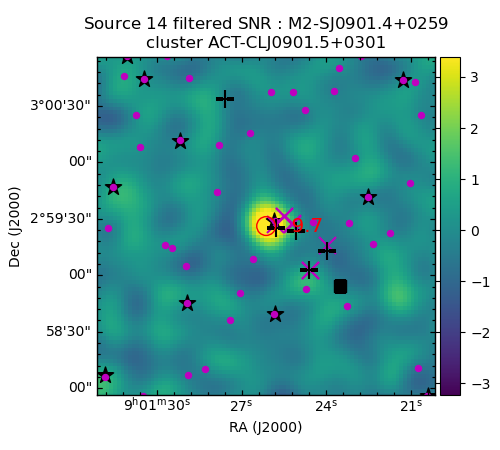}
    \includegraphics[width=0.23\linewidth]{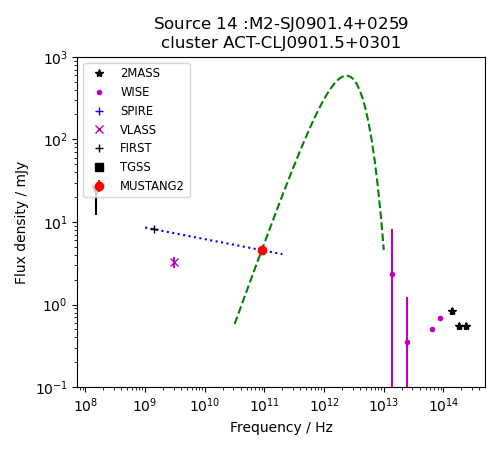}
    
    \includegraphics[width=0.23\linewidth]{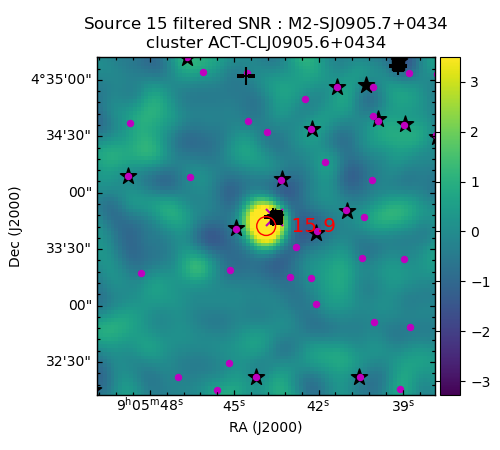}
    \includegraphics[width=0.23\linewidth]{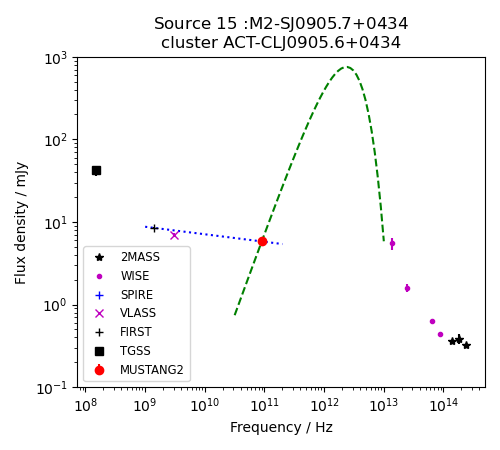}\hspace{0.8cm}
    \includegraphics[width=0.23\linewidth]{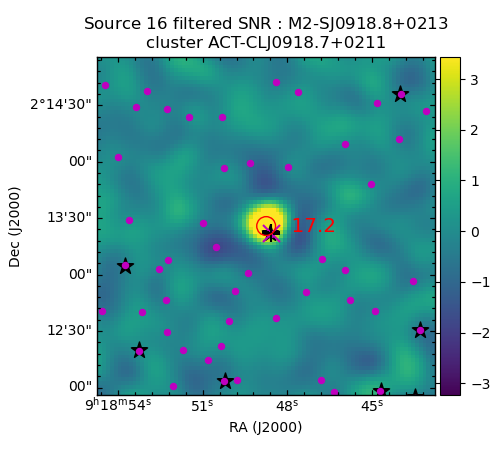}
    \includegraphics[width=0.23\linewidth]{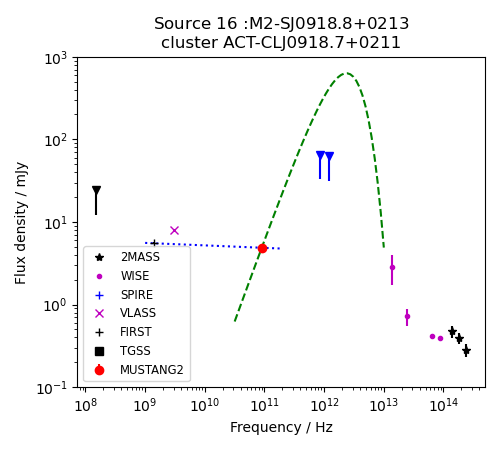}
    
    \includegraphics[width=0.23\linewidth]{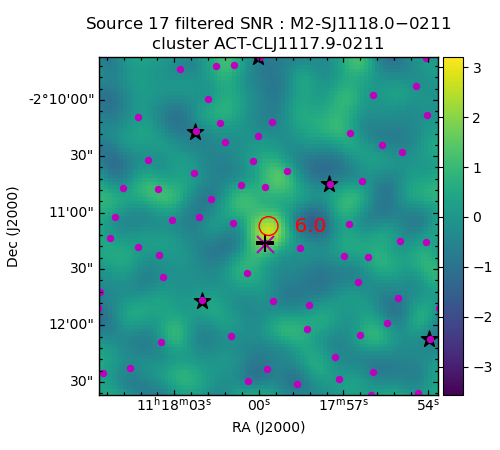}
    \includegraphics[width=0.23\linewidth]{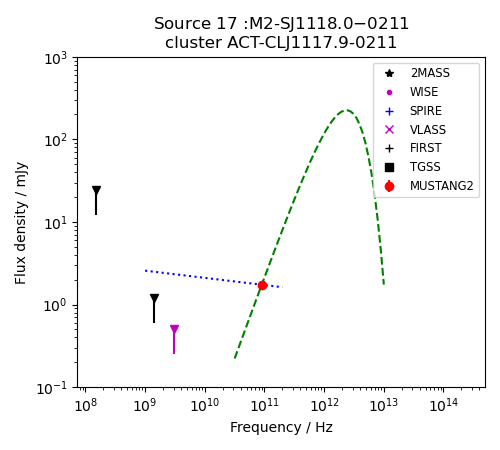}\hspace{0.8cm}
    \includegraphics[width=0.23\linewidth]{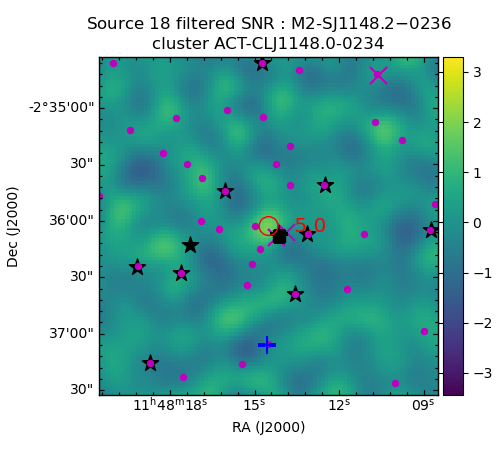}
    \includegraphics[width=0.23\linewidth]{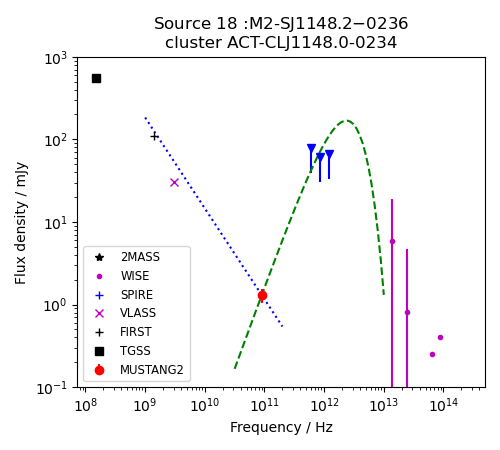}
    
   \includegraphics[width=0.23\linewidth]{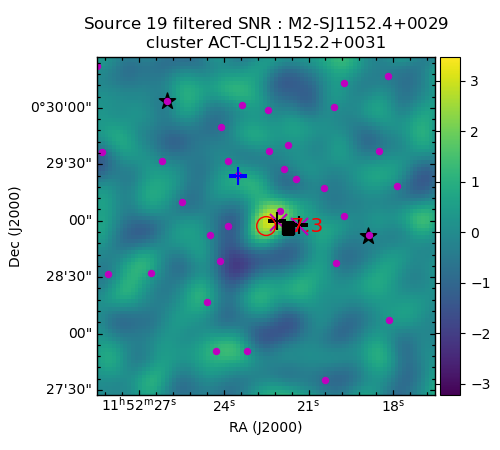}
    \includegraphics[width=0.23\linewidth]{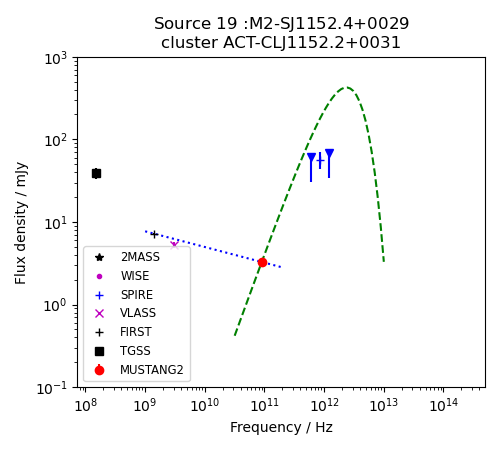}\hspace{0.8cm}
    \includegraphics[width=0.23\linewidth]{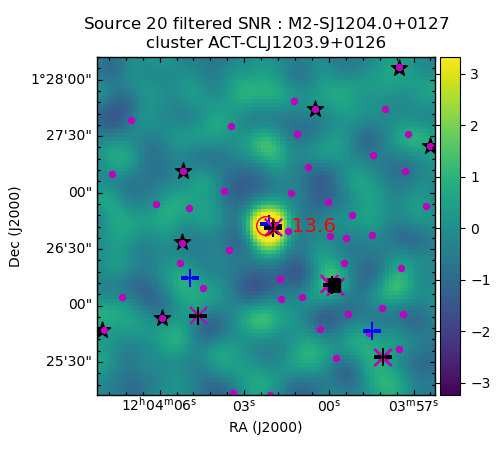}
    \includegraphics[width=0.23\linewidth]{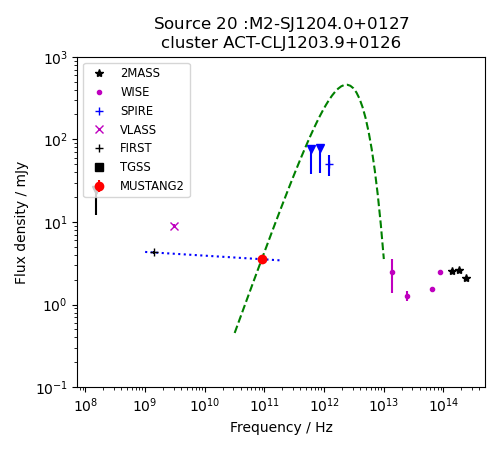}
    
    \caption{On-line Extra: \figfourcaption \newline
    }
\end{figure*}

\renewcommand{\thefigure}{4c}
\begin{figure*}
    \centering
   \includegraphics[width=0.23\linewidth]{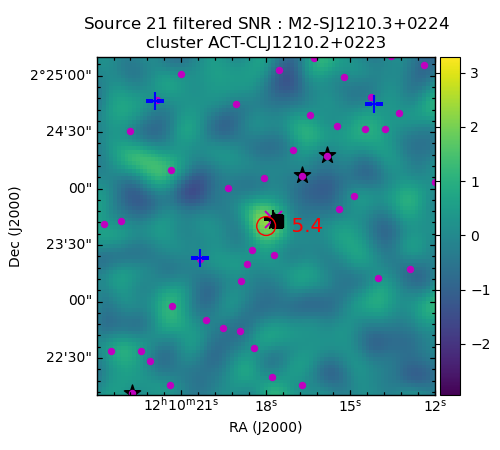}
    \includegraphics[width=0.23\linewidth]{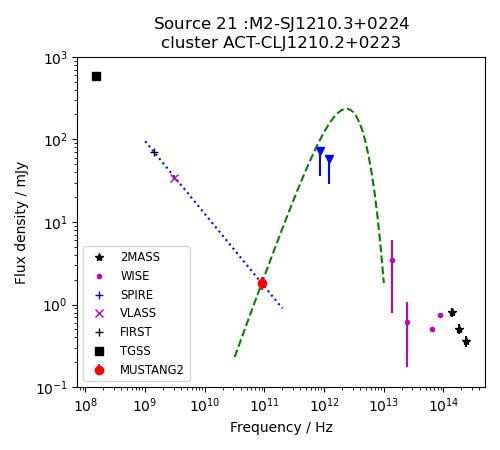}\hspace{0.8cm}
    \includegraphics[width=0.23\linewidth]{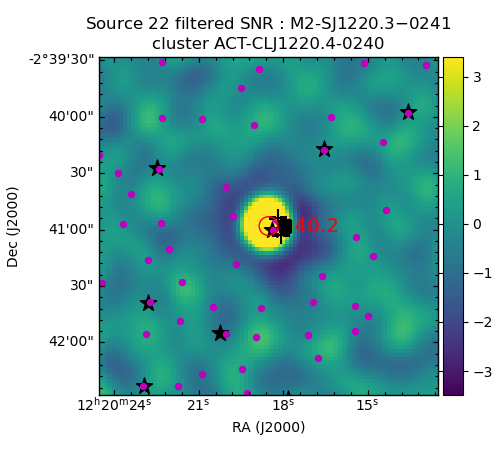}
    \includegraphics[width=0.23\linewidth]{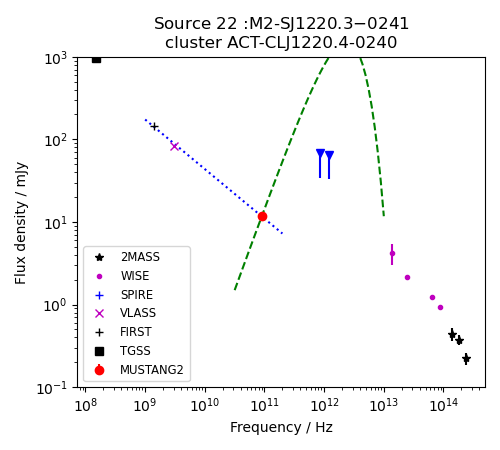}
    
   \includegraphics[width=0.23\linewidth]{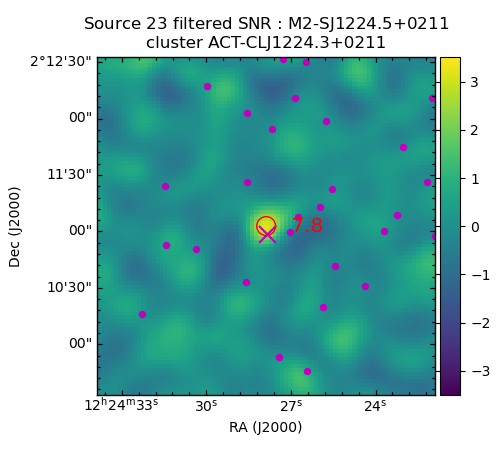}
    \includegraphics[width=0.23\linewidth]{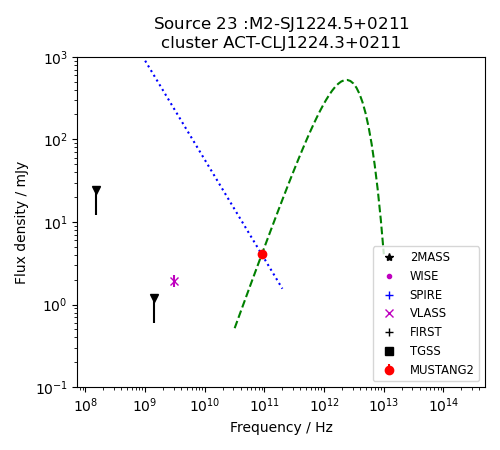}\hspace{0.8cm}
    \includegraphics[width=0.23\linewidth]{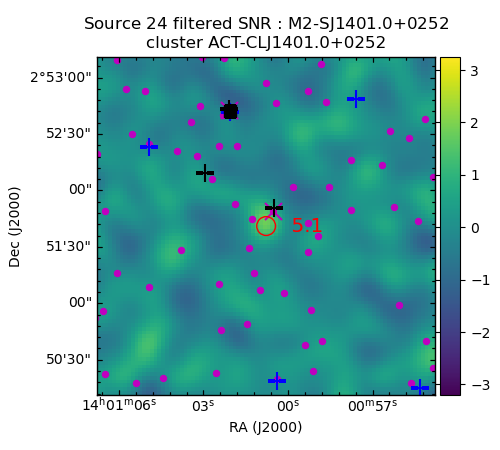}
    \includegraphics[width=0.23\linewidth]{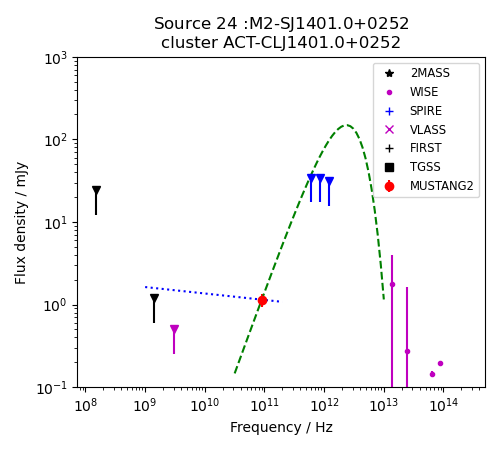}

    \caption{On-line Extra: \figfourcaption \newline
    }
\end{figure*}